\documentclass[
 reprint,
superscriptaddress,
 amsmath,amssymb,
 aps,
prstab,
nofootinbib
]{revtex4-1}

\usepackage{scrextend}
\usepackage{todonotes}
\usepackage{graphicx}
\usepackage{dcolumn}
\usepackage{bm}
\usepackage{amsmath}
\usepackage[colorlinks=true]{hyperref}
\usepackage{mathrsfs}
\usepackage{booktabs}

\begin{abstract}
    Linear $e^+e^-$ colliders enable precision measurements of Standard Model observables and beyond. Modern linear colliders need damping rings wherein the emittance of the source is damped and asymmetrized. A previous work [Phys. Rev. ST Accel. Beams 4, 021001 (2001)] delineated a systematic procedure for damping ring design. In this paper, we apply those principles, and frame damping ring design as a constrained numerical minimization problem when including intrabeam scattering. We use a multi-objective genetic optimization to map the transition from natural-emittance-dominated to intrabeam-scattering-dominated regimes. We compare to analytic results where possible to good agreement. The analysis is applied to parameters for the C$^3$ collider concept, but the procedure is fully generalizable to other linac parameters.
\end{abstract}

\begin{document}

\title{Damping Ring Optimization in Natural Emittance and IBS Dominated Regimes}

\author{M.B. Andorf} 
\affiliation{Cornell Laboratory for Accelerator Based Sciences and Education (CLASSE), Cornell University, Ithaca, NY 14853}
\author{N. Lockyer}
\affiliation{Cornell Laboratory for Accelerator Based Sciences and Education (CLASSE), Cornell University, Ithaca, NY 14853}
\author{E. Nanni} 
\affiliation{SLAC National Accelerator Laboratory, 2575 Sand Hill Road, Menlo Park, CA 94025}
\author{C. Vernieri}
\affiliation{SLAC National Accelerator Laboratory, 2575 Sand Hill Road, Menlo Park, CA 94025}
\author{J. Maxson}
\affiliation{Cornell Laboratory for Accelerator Based Sciences and Education (CLASSE), Cornell University, Ithaca, NY 14853}
\affiliation{Cornell High Energy Synchrotron Source (CHESS), Cornell University, Ithaca NY, 14853}

\maketitle
\section{Introduction}
A high energy $e^+e^-$ collider would enable precision measurements of the Higgs boson, the top quark, electroweak observables and has the potential to reveal new physics beyond the standard model~\cite{Narain:2022SnowmassEF,P5:2023Report,abramowicz2025linear,ESPPU:2026}.
A Linear-Collider (LC) facility with an energy scale of 100's of GeV  $e^{+}e^{-}$ collisions is an effective approach. One of the key components of an LC is Damping Rings (DRs) which reduces the normalized beam emittance from the electron/positron sources before injection into the main linac for acceleration  to the collision energy. A DR relies on synchrotron radiation (SR) for damping to produce a flat-beam ($\sigma_x \gg \sigma_y$) that is beneficial in reducing the beamstrahlung effect~\cite{Yokoya1992}.

Over the years several LCs have been proposed with each collaboration putting forth a mature DR design based on the collider parameters and luminosity goals of the research program~\cite{CLIC_cdr,ILC_cdr}.
In this work we study the parameters of the Cool Copper Collider ($C^3$)~\cite{vernieri2023cool,nanni2023status,Andorf:2025c3linear};  our approach is generalizable to other DR designs. We use a multi-objective genetic algorithm (MOGA) to design a DR based on C$^3$ parameters to optimize the horizontal beam emittance for the case when it is intrabeam scattering (IBS) dominated. Based on the systematic approach to DR design found in~\cite{emma2001systematic}, in this work we reframe DR design as a constrained optimization problem with the constraint arising from a maximum allowable damping time (set largely by the LC parameters) resulting in a minimum allowable beam energy and, therefore, achievable horizontal emittance. Analytic formulas are derived which allow for the zero-charge emittance to be optimized with a straightforward parameter scan. Lattice modeling is done in the \texttt{BMAD}-based Tool for Accelerator Optics (\texttt{TAO})~\cite{Sagan:Bmad,Sagan:TaoManual}, and a MOGA, implemented through \texttt{Xopt}~\cite{Xopt}, is used initially to verify analytic formulas before extending the modeling and optimization to include IBS.

\section{Optimized Natural Emittance of a DR}
For an LC the repetition rate of collisions, $f$, and bunch structure parameters are largely determined by the main linac and can be considered inputs for the DR design which results in constraints on the ring circumference and beam energy. The circumference of the ring is constrained by $C>N_{train}L_{train}$ where $L_{train}=(N_b-1)c\Delta t_b$ is the length of a bunch train with $N_b$ bunches spaced equally by $t_b$ and $N_{train}$ is the number of stored bunch trains in the ring.

The collision rate and number of stored bunch trains constrain the damping time as~\cite{emma2001systematic}
\begin{equation}
    \tau_y<\frac{N_{train}}{fN_{\tau}}.
    \label{tau_constrain}
\end{equation}
In the above, $N_{\tau}$ is the number of stored SR damping times between injection and extraction and the vertical damping time is used since the beam is injected round but extracted flat and therefore more damping times are needed in the vertical plane to reach equilibrium. As discussed in~\cite{emma2001systematic}, $N_{\tau}$ is a choice which is informed by balancing between too few damping times, in which case the extracted beam is significantly larger than the equilibrium emittance and too many damping times in which case the additional store time yields a negligible reduction in the extracted emittance and unnecessarily increases the energy of the DR which increases the natural emittance and makes magnet design more difficult. In this work we select $N_{\tau}=5.5$. 

The $C^3$ bunch structure varies depending on the operating mode. For our initial design, we use the baseline (BL) parameters given in Table~\ref{tab:paramsOperation}~\cite{Ntounis:2026beamBackgrounds}. We choose $N_{train}=4$ both to provide sufficient length for the damping wigglers and to accommodate the high-luminosity (high-$\mathscr{L}$) mode, in which the bunch train is doubled while the number of stored bunches is reduced to $N_{train}=2$. For the BL scenario, Eq.~\ref{tau_constrain} gives a maximum allowable vertical damping time of approximately $6$~ms. However, as shown below, when IBS is included in the equilibrium calculation, a damping time below approximately $3$~ms is favored. Thus, although the DR is designed using the BL parameters, the resulting design also satisfies the more demanding high-$\mathscr{L}$ operating scenario. The sustainability update (s.u.) requires a maximum vertical damping time below 12 ms and is thus also covered in the BL design. 
\begin{table}[t]
    \centering
    \caption{$C^3$ operating scenarios at a center-of-mass energy of 250~GeV:
    baseline (BL), sustainability update (s.u.), and high-luminosity
    (high-$\mathscr{L}$) modes.}
    \label{tab:paramsOperation}
    \renewcommand{\arraystretch}{1.2}
    \begin{tabular}{|l|cccccc|}
        \hline
        Mode & $N_b$ & $t_b$ (ns) & $L_{\rm train}$ (m)
        & $N_{\rm train}$ & $f$ (Hz) &  $\tau_y^{max}$ (ms) \\
        \hline
        BL & 133 & 5.26 & 208 & 4 & 120 & 6 \\
        \hline
        s.u. & 266 & 2.63 & 208 & 4 & 60 & 12 \\
        \hline
        high-$\mathscr{L}$ & 532 & 2.63 & 416 & 2 & 120 & 3 \\
        \hline
    \end{tabular}
\end{table}

The SR damping time in the transverse plane is given by~\cite{Wolski:2015lowemittance,Wiedemann:2007particle}
\begin{equation}
\tau_{i}=\frac{2EC}{j_{i}U_o c}
\end{equation}
where E is the beam energy, $i=x,y$ for either the horizontal or vertical plane and
\begin{equation}
U_o=\frac{C_\gamma E^4 I_2}{2\pi}
\end{equation}
is the energy loss due to SR over one turn of the ring, $C_{\gamma}\approx8.846\times10^{-5}$ m/GeV$^3$ and $I_2$ is the second SR integral $I_2=\oint \frac{ds}{\rho^2}$ where the integration is over the circumference of the ring and $\rho(s)$ is the local bend radius.  $j_x=1-I_{4}/I_{2}$ is the partition number where $I_{4}=\oint \frac{\eta_x}{\rho}(\frac{1}{\rho^2}+2k_1)$, $\eta_x$ is the horizontal lattice dispersion and $k_1=\frac{e}{P_o}\frac{\partial B_y}{\partial x}$. By design $\eta_y=0$ so $j_y=1$. 

Since damping time depends on the beam energy Eq.~\ref{tau_constrain} leads to a constraint on its minimum value:
\begin{equation}
E_{min}=\bigg(\frac{4\pi L_{train} N_{\tau}f}{C_{\gamma} c I_2}\bigg)^{1/3}
\label{Emin}
\end{equation}
where we set $C=N_{train}L_{train}$. The minimum energy is independent of the choice of stored bunch trains since increasing, for example, the stored bunch trains from one to two doubles the circumference and consequently the damping time but the constraint from Eq.~\ref{tau_constrain} also doubles. Therefore, the choice of number of bunch trains is only a matter of balancing space in the ring for beamline components and keeping the overall ring size down. The normalized horizontal natural equilibrium emittance from SR is
\begin{equation}
\epsilon_x=C_q\gamma^3\frac{I_5}{j_x I_2}
\label{em_gen}
\end{equation}
where $C_q\approx 3.8319 \times 10^{-13}$ m and $I_{5}=\oint \frac{\mathcal{H}ds}{\mid{\rho^3}\mid}$. $\mathcal{H}=\gamma\eta^2+2\eta\eta'\alpha+\beta\eta^{'2}$ is the dispersion invariant. Taken together Eq's \ref{Emin} and \ref{em_gen} indicate that to optimize the natural emittance, the beam energy should be set to $E_{min}$ and the obtainable horizontal emittance scales linearly with the bunch train length and collider repetition rate. 

In this work we consider a race-track layout where the ring consists of two arcs and two-straight sections. In the arcs, periodic Theoretical Minimum Emittance (TME) cells are selected since they provide the lowest possible contribution to emittance growth in the bending magnets and are also relatively simple in layout. The emittance scales with the number of bending magnets, $N_d$ as
\begin{equation}
    \epsilon_{ring}=FC_q\gamma^3\frac{8\pi^3}{N_d^3}
    \label{em_nd}
\end{equation}
where $F$ is a numerical constant that depends on the design of the arc-cells. For an optimally tuned TME cell $F_{TME}=\frac{1}{12\sqrt{15}}$.  An optimally tuned TME cell requires strong focusing resulting in high chromaticity; therefore, TME cells are often detuned and $F$ will be a factor of order unity larger than $F_{TME}$. For the case of identical dipoles of length $L_d$ in the arcs 
\begin{equation}
I_{2,ring}=\frac{4\pi^2}{N_dL_d}. 
\label{I2_arc}
\end{equation}

Additionally, by equating Eq.'s \ref{em_gen} and \ref{em_nd} we obtain 
\begin{equation}
I_{5,ring}=\frac{2^5\pi^5Fj_x}{N_d^4L_d}.
\label{I5_arc}
\end{equation}

The straight sections of the layout create space for dispersion free sections where damping wigglers can be installed. Damping wigglers reduce the beam emittance in two ways: the additional bending in the wiggler contributes to the $I_2$ integral directly reducing the equilibrium emittance as evident in Eq.~\ref{em_gen}. The additional damping from the wiggler also lowers the minimum required energy which further reduces the equilibrium emittance.

We model the damping wiggler as a sequence of dipoles of alternating polarity and constant bending radius $\rho_{wig}$  yielding:
\begin{equation}
I_2=L_{wig}/\rho_{wig}^2.
\label{I2_wig}
\end{equation}
For simplicity, this idealized model neglects the half-strength end poles required for dispersion closure at the wiggler boundaries. 

The dispersion that arises within the wiggler contributes to $I_{5}$ which becomes non-negligible for large field strengths. The wiggler period, $\lambda_{wig}$ is twice the length of a single dipole which has a bending angle $\theta_{wig}=\lambda_{wig}/2\rho_{wig}$.  Within the region of the full-strength dipoles the dispersion function is

\begin{equation}
\begin{aligned}
\eta_{\mathrm{wig}}(s)
&= \rho_{\mathrm{wig}}\!\left(1-\cos\bigg(\frac{\theta_{\mathrm{wig}}}{2}\bigg)\right) \\
&\quad + 2\rho_{\mathrm{wig}}\sin\!\left(\frac{\theta_{\mathrm{wig}}}{2}\right)
\tan\!\left(\frac{\theta_{\mathrm{wig}}}{4}\right)\cos(k_{\mathrm{wig}} s)
\end{aligned}
\end{equation}

which is a sinusoidal function in $s$ with a constant offset.  Inside the wiggler, the focusing is weak such that the Courant-Snyder parameter $\alpha\approx0$ and consequently $\gamma\approx1/\beta$. Moreover, since dispersion remains small in the wiggler, $\left| \eta \right| \ll \beta \left| \eta'\right|$, the dispersion invariant is approximately $\mathcal{H}\approx \eta'^{2}_{wig}\beta$~\cite{Wiedemann:2007particle} so that
\begin{equation}
I_{5,wig}\approx\oint\frac{\eta'^{2}_{wig}\beta_x ds}{\rho_{wig}^3}\approx \frac{L_{wig}\pi^2\lambda_{wig}^2\langle \beta_x\rangle}{2^9\rho_{wig}^5}
\label{I5_wig}
\end{equation}
where, consistent with $\alpha\approx 0$, to evaluate the integral $\beta$ is approximated as constant and replaced with the mean value in the wiggler. In the above we also used
\begin{equation}
\sin{(\theta_{wig}/2)}\tan{(\theta_{wig}/4)}\approx\frac{\theta_{wig}^2}{8}=\frac{\lambda_{wig}^2}{16 \rho_{wig}^2}.
\end{equation}
since the bending angle in the wiggler can be assumed small.

The total emittance of the DR can now be expressed as
\begin{equation}
\begin{aligned}
\epsilon= &C_q\gamma^3\frac{I_{5,ring}+I_{5,wig}}{I_{2,ring}+I_{2,wig}}\\=
&\frac{\mathcal{M}}{j_x}\frac{I_{5,ring}+I_{5,wig}}{(I_{2,ring}+I_{2,wig})^2}\\ =
&\frac{\mathcal{M}}{j_x}\frac{\frac{2^5\pi^5 j_xF}{N_d^4L_d}+\frac{L_{wig}\pi^2\lambda_{wig}^2\langle \beta_x\rangle}{2^9\rho_{wig}^5}}{(\frac{4\pi^2}{N_d L_d}+\frac{L_{wig}}{\rho_{wig}^2})^2}.
\end{aligned}
\label{em_total}
\end{equation}
where we have set the beam energy to the minimum energy given by Eq.~\ref{Emin}
and defined a constant $\mathcal{M}=\frac{C_q}{C_\gamma}\frac{4\pi L_{train}N_{\tau}f}{c (m_ec^2)^3}$.  Notice that $\mathcal{M}$ contains only universal constants and parameters that are specific to a particular LC, while the radiation integrals that appear above are generic to any DR (e.g. not specific to a particular LC). Therefore, $\mathcal{M}$ shows how the natural emittance scales with LC parameters.
In the above, we set $j_x\approx1$ since the DRs considered in this work have damping dominated by the wiggler and $I_{4,wig}/I_{2,wig}\approx0$~\cite{Jang2020}.

To reduce the natural normalized horizontal emittance, the condition for a damping wiggler of any length is  (see Appendix \ref{Appendix A})
\begin{equation}
\epsilon_{wig}<2\epsilon_{ring}
\end{equation}
where $\epsilon_{wig}$ is the emittance purely from the wiggler and given by Eq.~\ref{em_wig} and $\epsilon_{ring}$ is the emittance of the arcs. In general, for a DR there is no finite value of $L_{wig}$ that minimizes the natural emittance. The emittance always tends to 0 as the wiggler length is increased to infinity since $\gamma$ tends to zero. If the above inequality is not met a global   \textit{maximum}  exists at a non-zero length. The wiggler period only appears in the numerator of $I_{5,wig}$ which implies a smaller wiggler period is always favored for lower emittance at a given field strength. For a given value of $L_{wig}$ and $\lambda_{wig}$ there is an optimal value of $\rho_{wig}$ that is easily found numerically.

\begin{figure}
    \centering
    \includegraphics[width=1\linewidth]{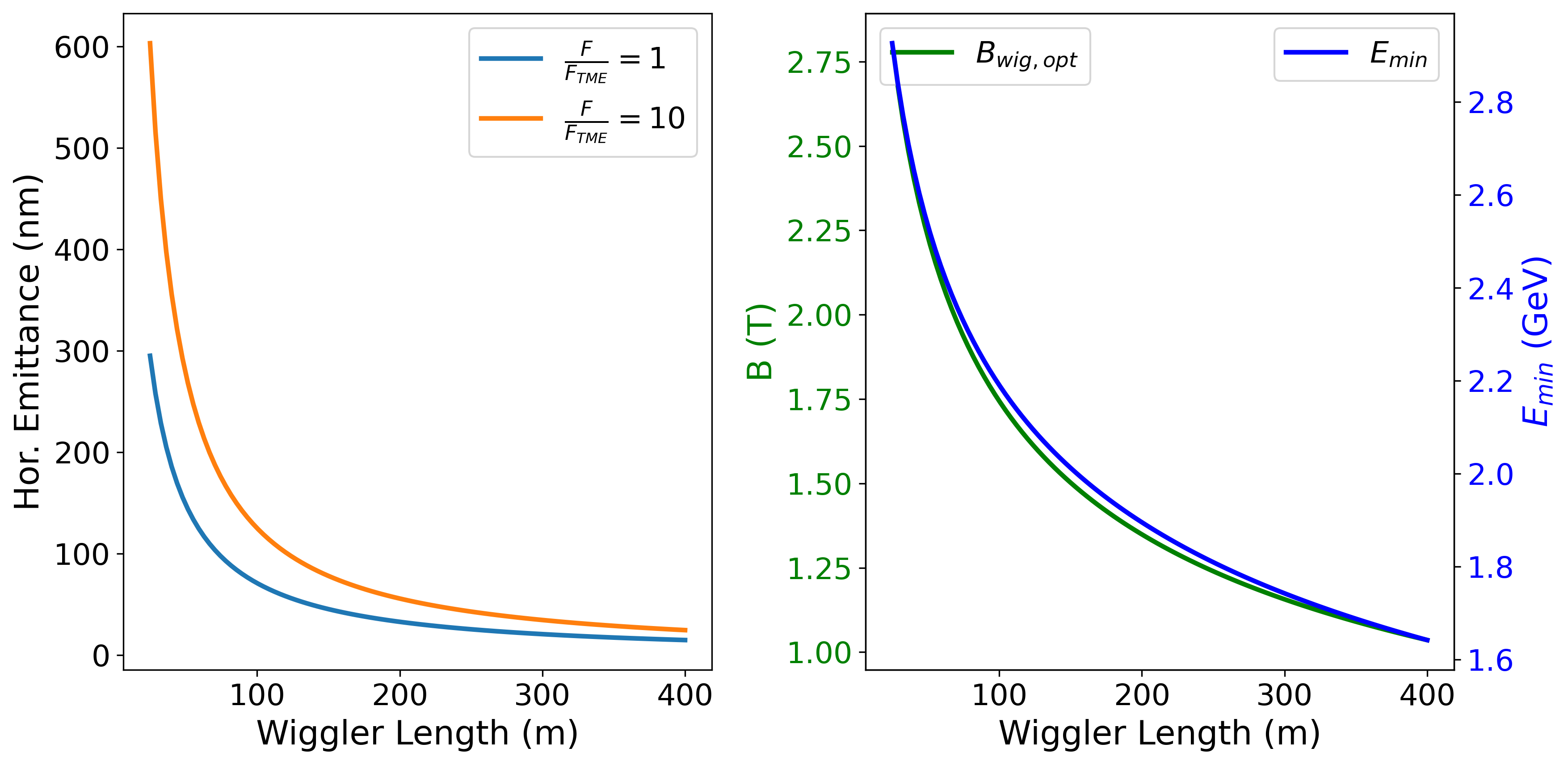}
    \caption{(Left)  optimal emittance as a function of wiggler length. (Right) optimal B field and beam energy as a function of wiggler length.}
    \label{fig:analy1}
\end{figure}
In the left panel of Fig.~\ref{fig:analy1}, the optimal emittance as a function of wiggler length is computed assuming the parameters found in Table ~\ref{tab:paramsOperation} for BL scenario and in Table~\ref{tab:params}. The arc-cells are assumed to be TME, and two cases are considered: one in which the arc-cells are optimally tuned to $F=F_{tme}$ and another where they are detuned to $F=10F_{tme}$. In addition to reducing the emittance, the use of a damping wiggler relaxes the need for the arc-cells to be optimally tuned, allowing flexibility in their design if chromaticity is an issue. In the right panel, the optimal magnetic field for the case of $F=F_{tme}$ and the resulting energy are shown.
\begin{table}
    \centering
    \begin{tabular}{l l l}\hline
         Parameter&  Value& Unit\\\hline
         $N_{\tau}$&  5.5& -\\
         $N_{d}$& 100&-\\ 
         $L_d$ & 0.7 & m\\ 
         $\lambda_{wig}$ &  4.9& cm \\ 
         $\langle \beta_x \rangle$& 3.0 & m \\ \hline
    \end{tabular}
    \caption{Assumed parameters for computations and optimizations of the natural emittance.}
    \label{tab:params}
\end{table}
\subsection{Natural Energy Spread}
The natural energy spread in a ring is~\cite{Wolski:2015lowemittance,Wiedemann:2007particle}:
\begin{equation}
\sigma_{p}^2=\gamma^2C_q\frac{I_3}{j_zI_2}
\end{equation}
where $j_z=2+I_4/I_2$ is the longitudinal partition number and 
\begin{equation}
I_3=\oint\frac{ds}{|\rho^3|}.
\end{equation}
The contributions to $I_3$ from bending magnets in the arcs and wigglers are
\begin{equation}
I_{3,arc}=\frac{8\pi^3}{N_d^2L_d^2}
\end{equation}
and 
\begin{equation}
I_{3,wig}=\frac{L_{wig}}{\rho_{wig}^3}
\end{equation}
and therefore at the minimum energy
\begin{equation}
\begin{aligned}
\sigma_{p}^2=&\frac{\mathcal{M}^{2/3}C_q^{1/3}}{j_z}\frac{I_3}{I_2^{5/3}}\\
=& \frac{\mathcal{M}^{2/3}C_q^{1/3}}{j_z}\frac{\frac{8\pi^3}{N_d^2L_d^2}+\frac{L_{wig}}{\rho_{wig}^3}}{\bigg(\frac{4\pi^2}{N_dL_d}+\frac{L_{wig}}{\rho_{wig}^2}\bigg)^{5/3}}.
\end{aligned}
\end{equation}
\begin{figure}
    \centering
    \includegraphics[width=0.75\linewidth]{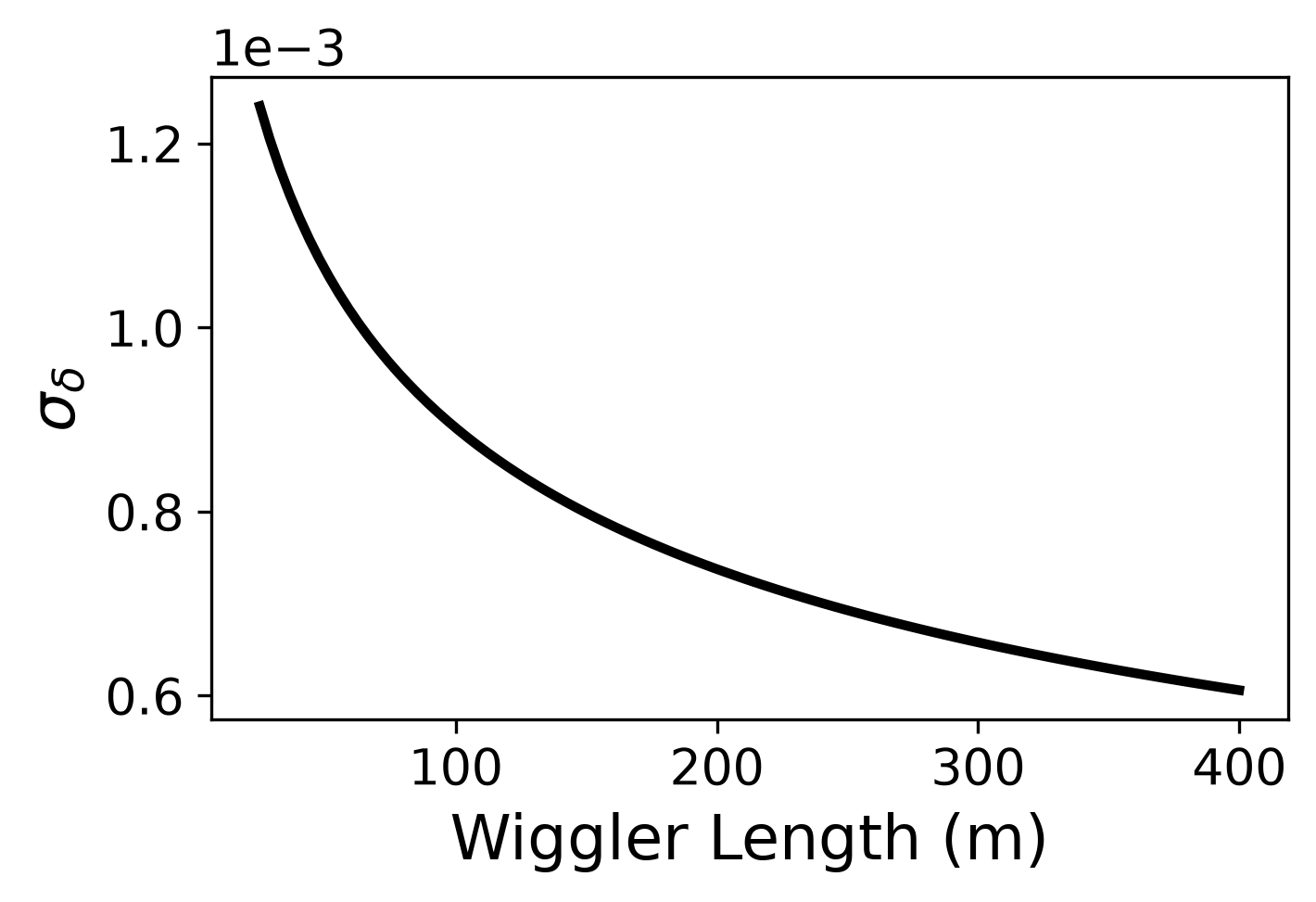}
    \caption{The resulting natural energy spread at optimal emittance.}
    \label{fig:analyticEnergy}
\end{figure}
The resulting natural energy spread at the optimal emittance is shown in Fig.~\ref{fig:analyticEnergy}, where $j_z=2$ has been assumed. Note that the relative energy spread scales as $(fL_{\rm train})^{1/3}$, whereas the normalized horizontal emittance scales linearly.

After extraction from the DR, the beam is compressed longitudinally before being boosted to collision energy. In the case of $C^3$, the bunch will be compressed to approximately 100 $\mu$m. In the DR, the bunch length is set by the amplitude of the RF cavities and a bunch length of order millimeters is typical. During bunch compression, neglecting any effect of Coherent Synchrotron Radiation (CSR), the longitudinal emittance is conserved, $\epsilon_z=\sigma_z\sigma_{\delta}$ where before and after compression there is no correlation between particle position and energy. Consequently, a relative energy spread of order 1\% can be expected after compression. IBS growth inside the DR will cause this number to grow and will be computed later on.

\section{MOGA Optimization of Natural Emittance}
In this section, we present results from a multi-objective genetic algorithm (MOGA) study. The optimization both validates the analytic predictions developed in the previous section and establishes a numerical framework based on lattice modeling in TAO that is extended to include IBS in the following section. The optimization targets the natural emittance and total wiggler length as objectives. Since the analytic theory predicts no finite optimum in wiggler length, practical considerations, including cost and technical complexity, motivate treating the wiggler length as a competing objective and examining the resulting trade-off.

 A full ring lattice is not constructed during the optimization. Instead, for each lattice setting, a single arc-cell and wiggler cell is made. We then use the fact that a full ring is comprised mostly of a definite number of arc-cells (100) and wiggler cells (dependent on the total wiggler length) to compute the radiation integrals. This approach does not include a section between the two cell types for beta-function matching and dispersion suppression but allows for rapid scanning of parameters. As a check to this composite approach, a full ring based on one of the MOGA individuals was constructed and the resulting lattice and beam parameters were found to be in good agreement to about $1\%$ (e.g. the matching section makes a negligible impact). 
 
The arc-cells have the form of a TME cell with two focusing quadrupoles flanking a single bend magnet with a defocusing gradient. While chromaticity is not considered in the MOGA, space is allocated for two families of sextupoles. One family is situated at the start and end of the cell while the other is between the focusing quads and bend magnet. This arrangement ensures chromatic corrections are possible and is similar to the design used in CLIC~\cite{CLIC_cdr}.  The MOGA is allowed to vary the strengths and locations of the focusing quads and the strength of the dipole gradient. The number of arc-cells is fixed to 100. The length of the arc-cell for each individual is varied based on the MOGA's selected wiggler length such that the ring circumference $C=N_{train}L_{train}$ stays fixed for all individuals. The wiggler cells are FODO cells with the wiggler filling space between the quadrupoles. 
To reduce the emittance contribution from the wiggler, the average horizontal beta function $\langle \beta_x \rangle$ should be minimized. For a FODO cell of total length $L_{\mathrm{FODO}}$, the minimum achievable value is $\langle \beta_x \rangle_{\min} = L_{\mathrm{FODO}}/2$. This condition is not enforced explicitly in the MOGA, since the quadrupole strengths
required to reach $\langle \beta_x \rangle_{\min}$ depend on the wiggler field strength,
which introduces additional focusing. Instead, the quadrupole strengths in the wiggler
cells are treated as free variables in the optimization. In practice, the selected
solutions typically yield a mean beta function close to $\langle \beta_x \rangle_{\min}$. 

Although it is clear that for the case of optimizing the natural emittance the beam energy should be set to the minimum value given by Eq.~\ref{Emin}, the energy is treated as variable and $E_{min}$ is incorporated as a constraint $E<E_{min}$ evaluated for each individual after the wiggler parameters have been selected. A list of variables for the MOGA is shown in Table~\ref{tab:params2}. Additionally, the location of the quadrupole in the TME cell was allowed to vary freely from the start of the cell to the start of the bend.

\begin{figure}
    \centering
    \includegraphics[width=1\linewidth]{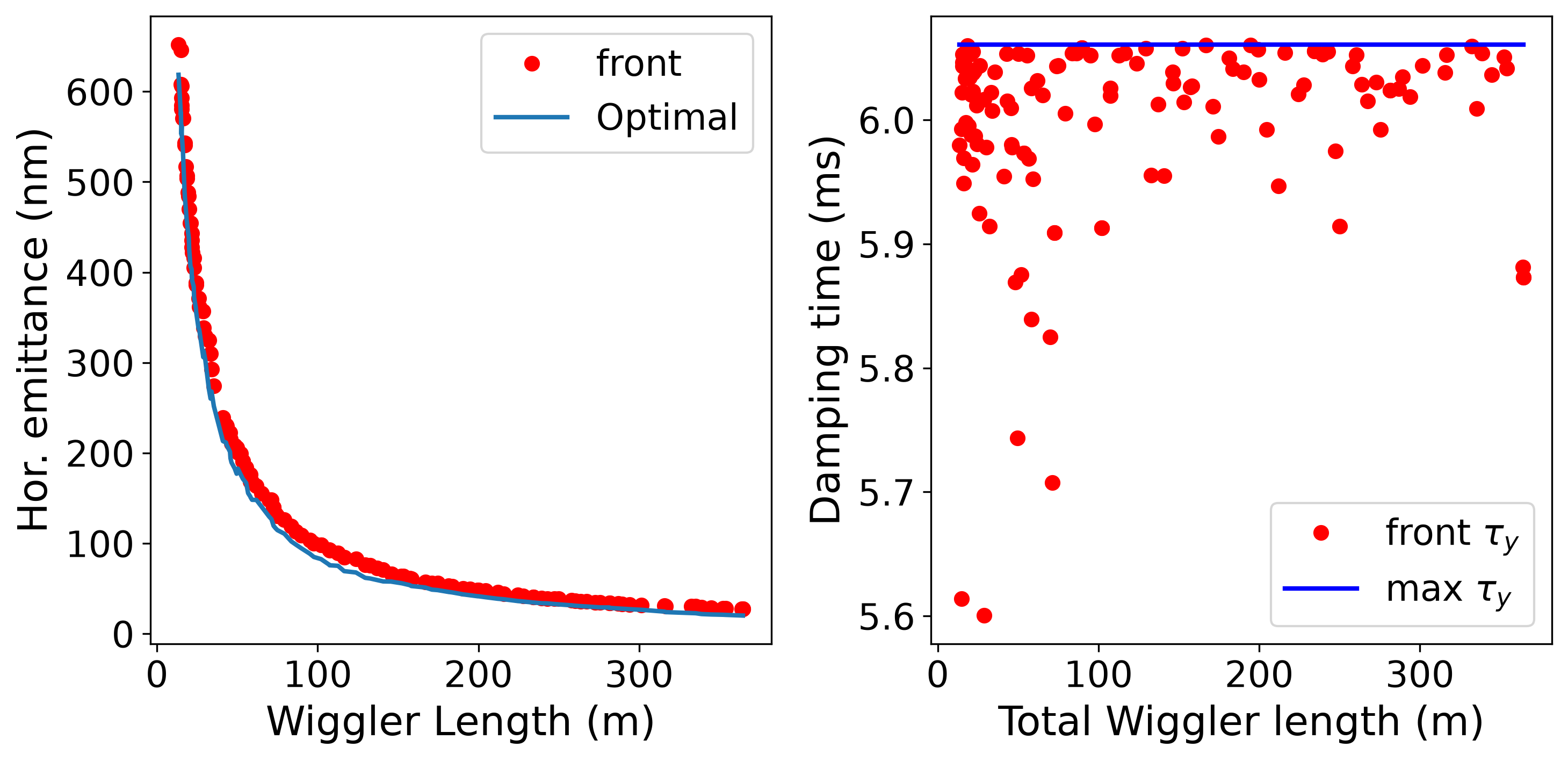}
    \caption{ (Left) The resultant Pareto front of the MOGA (red dots) and optimal emittance computed from Eq.~\ref{em_total}. (Right) the resulting vertical damping time of each individual (red dots) and the maximum allowable damping time derived from Eq.~\ref{tau_constrain}}
    \label{fig:natural_front}
\end{figure}

\begin{figure}
    \centering
    \includegraphics[width=1\linewidth]{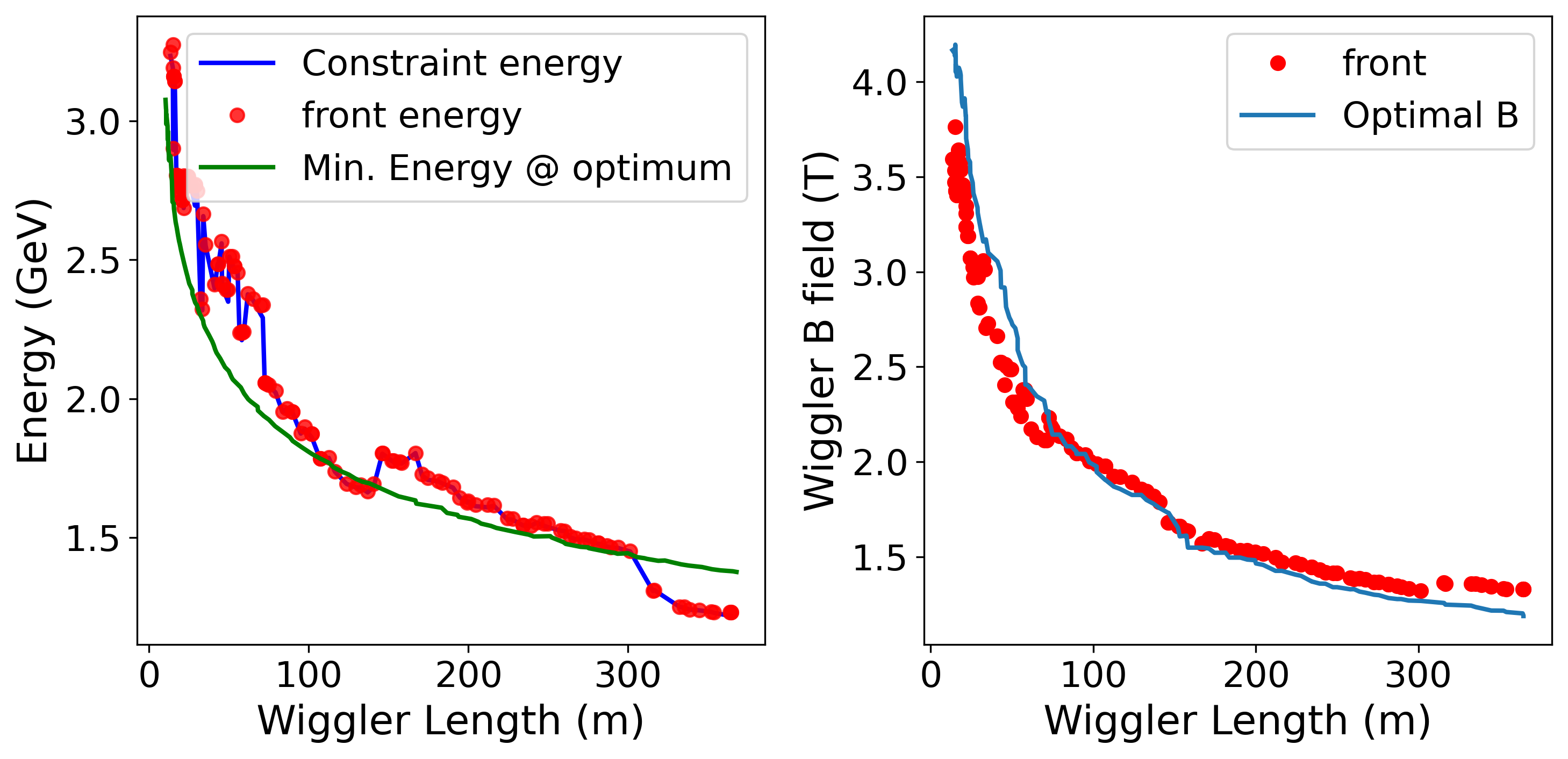}
    \caption{(Left) The energy as a function of $L_{wig}$ for individuals on the Pareto front (red dots), the constraint computed from Eq.~\ref{Emin} (blue line) and the optimal value determined by Eq.~\ref{em_total}.  (Right) The wiggler magnetic field as a function of its length for (red dots) individuals along the Pareto front and (blue line) the optimal value found with Eq.~\ref{em_total}.}
    \label{fig:Natural Energy and Damping}
\end{figure}

\begin{table}
    \centering
    \begin{tabular}{l l l}\hline
         Variable &  Range & Unit\\\hline
         $K_1^{Q}$ (TME) &  (0,6) & m$^{-2}$\\
         $K_1^{B}$ (TME) & (-1.5,0) & m$^{-2}$\\
         Wiggler Length & (5,370) & m\\
         $K_1^{Q}$ (FODO) & (-0.9,2.3) & m$^{-2}$\\
         Wiggler Field & (0,4) & T\\ 
         Beam Energy & (1.0,6.0) & GeV\\ \hline
    \end{tabular}
    \caption{Assumed parameters for computations and optimizations of the natural emittance.}
    \label{tab:params2}
\end{table}
The resulting approximate Pareto front is shown in Fig.~\ref{fig:natural_front}. Individuals are plotted in red dots and are seen to be in good agreement with the optimal front using the approach developed in the previous section plotted in blue. In the right panel of Fig.~\ref{fig:natural_front}, the resulting vertical damping time is computed and compared to the maximum allowable damping time given by Eq.~\ref{tau_constrain} and plotted in blue. Individuals are clustered near the maximum allowable damping time but no individual exceeds it. This is a direct consequence of the energy constraint as the optimizer is not directly computing the damping time in the optimization.  The energies of the individuals on the approximate front are shown in the left panel of Fig.~\ref{fig:Natural Energy and Damping} as red dots. For each individual's $I_2$ the resulting $E_{min}$ is plotted in blue while $E_{min}$ for the optimal wiggler strength computed from Eq.~\ref{em_total} is shown in green.  The energy of each individual closely matches the imposed energy constraint, but deviates slightly from the analytic optimum. To understand the origin of these deviations, the right panel compares the wiggler magnetic field strength of the MOGA-selected solutions to the optimal value. We see that when solutions with field strengths exceeding the analytic optimum are selected, they are compensated by a corresponding reduction in beam energy, as permitted by Eq.~\ref{Emin} and no corresponding deviation between the analytically predicted and MOGA-derived value in the horizontal emittance is observed. This can be further understood by noting that solutions near the front tend to be wiggler dominated so that $\epsilon\approx \epsilon_{wig} $ which goes as both $\propto\gamma^3$ and $\approx1/\rho_{wig}^3\propto B_{wig}^3$. (see Eq.~\ref{em_wig}). Therefore, we would expect the gradient with respect to these parameters to be relatively flat near the optimal and the MOGA will struggle to converge onto the true optimal solution.

\section{MOGA Optimization with IBS}

\begin{figure}
    \centering
    \includegraphics[width=1\linewidth]{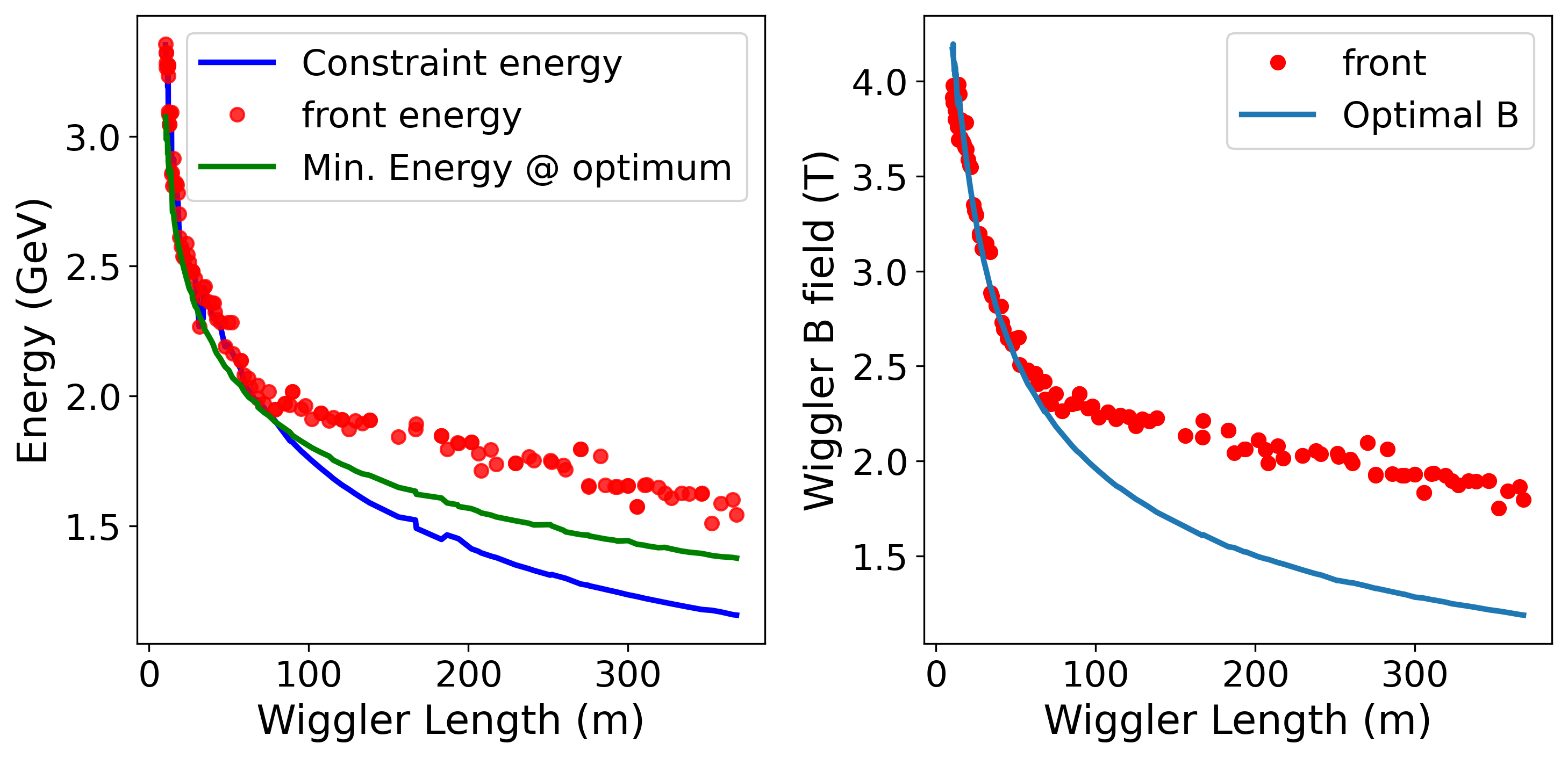}
    \caption{(Left) The energy for individuals on the Pareto front (red dots), the constraint computed from Eq.~\ref{Emin} (blue line) and the optimal value determined by Eq.~\ref{em_total}.  (Right) The wiggler magnetic field for (red dots) individuals along the Pareto front and (blue line) the optimal value found with Eq.~\ref{em_total}.}
    \label{fig:IBS Energy and Damping}
\end{figure}

The numerical model developed in the previous section can be extended to include IBS.  To compute the growth rates in transverse emittances and longitudinal energy spread we use the Completely Integrated Modified Piwinski (CIMP) formulation~\cite{KUBO_CIMP}.  Details of its implementation can be found in Appendix~\ref{Appendix B}.
\begin{figure}
    \centering
    \includegraphics[width=1\linewidth]{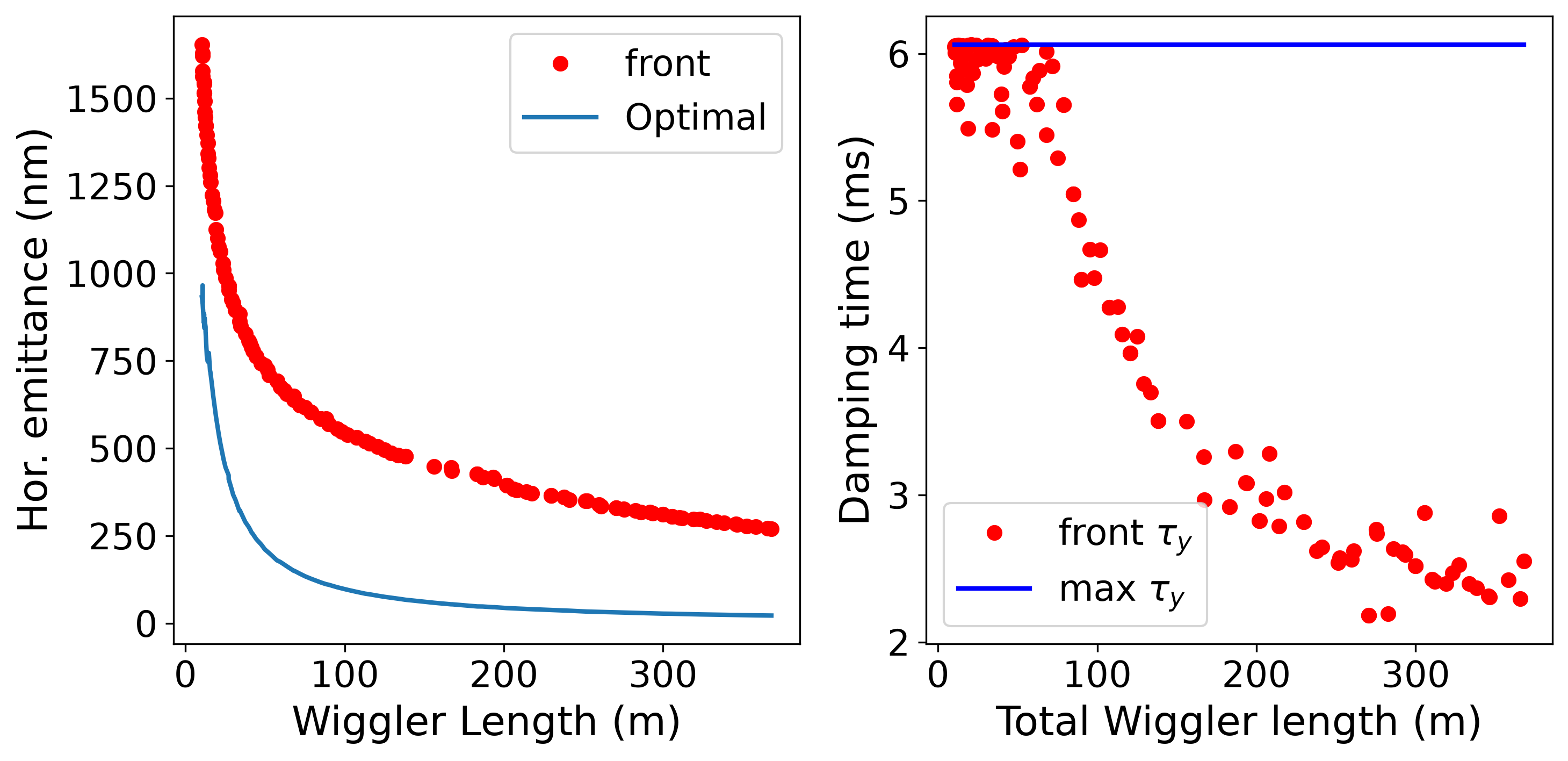}
    \caption{(Left) The resultant approximate Pareto front of the MOGA (red dots), accounting for IBS, and the optimal (natural) emittance values in blue. (Right) The vertical damping time of individuals on the approximate Pareto front. }
    \label{fig:IBSFront}
\end{figure}
The MOGA described in the previous section was re-run with IBS included, for a bunch charge of 1 nC. Since the IBS rates depend  on the bunch length, an RF element is included and its voltage is automatically adjusted to maintain a 2-mm natural bunch length based on the lattice settings of each individual. The resulting approximate Pareto front is shown in red in Fig.~\ref{fig:IBSFront} and the optimal natural emittance is shown in blue. As expected, the equilibrium emittance is larger when IBS is included. The right side of Fig.~\ref{fig:IBSFront} shows the vertical damping time of individuals on the front. Individuals with a wiggler length approximately greater than 75 m have damping times significantly less than the constraint imposed from Eq.~\ref{tau_constrain}. so much so that for total wiggler lengths beyond approximately 200 meters, the damping time is sufficiently fast to handle the $N_{train}$=2 case (e.g. $\tau_y<$3 ms)  selected for the high luminosity operating mode of $C^3$ without needing to modify the settings of the DR. Furthermore, it can be seen in Fig.~\ref{fig:IBS Energy and Damping} that for individuals with a wiggler length less than approximately 75 m, the MOGA selects an energy and wiggler B-field that closely matches the analytic treatment from the previous section while a sharp deviation from the analytic values can be seen for individuals with a wiggler length greater than approximately 75 m where both higher energy and magnetic fields are favored.

By comparing the blue and red curves in the left panel of Fig.~\ref{fig:IBSFront}, we see the deviation occurs when the inclusion of IBS results in approximately a doubling of the emittance compared to the natural value. Therefore, there are two regimes: an SR excitation dominated regime where the majority of the emittance is the result of photon emission in a dispersive region of the ring and an IBS-dominated regime. In the SR excitation regime, even though the increase in emittance due to IBS can be substantial (up to a factor of 2), additional damping does not help because it results in larger emittance growth from SR which is the dominating factor. In the IBS-dominated regime, additional damping, despite increasing growth from SR excitation, helps because IBS is the dominating factor. Additionally, in this regime, the $1/\gamma^4$ scaling of the IBS rate is leveraged by increasing the beam energy beyond the constraint imposed by the damping time.
\begin{figure}
    \centering
    \includegraphics[width=0.75\linewidth]{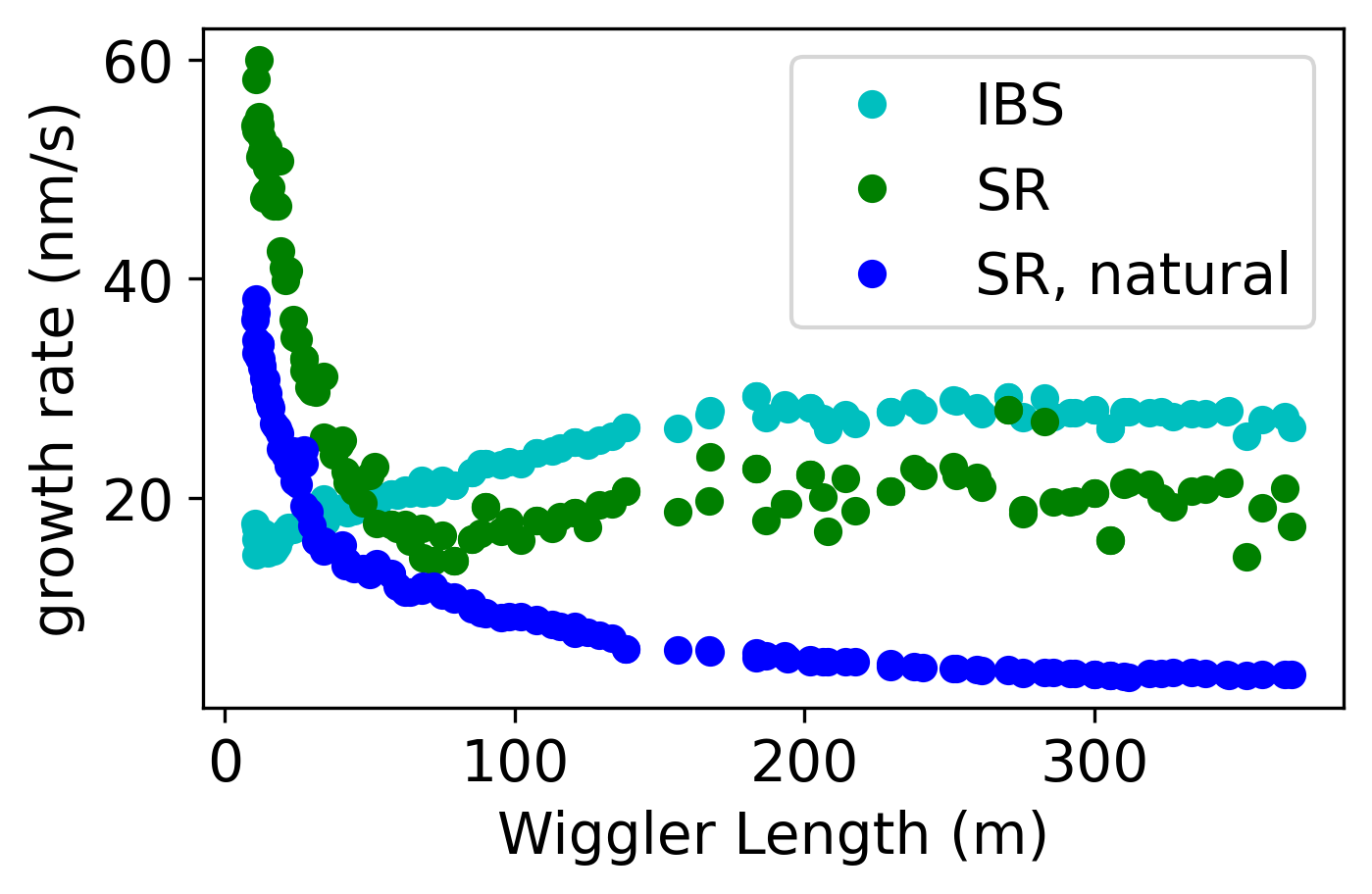}
    \caption{Emittance growth rates from IBS (light blue) and SR excitation (green) for the MOGA optimization accounting for IBS. For comparison, the resulting SR excitation growth rate for the natural emittance MOGA is shown in dark blue. }
    \label{fig:Growth rates}
\end{figure}

The regime behavior can be further illuminated by comparing the growth rates from SR and IBS at equilibrium as shown in Fig.~\ref{fig:Growth rates}.  For the natural emittance case (dark blue), as the wiggler length is increased damping is achieved with monotonically decreasing SR-excitation resulting in the previously observed decrease in the emittance. For the case when IBS is included, initially we observe a decrease in SR-excitation (green) while the IBS rate (light blue)--which is initially smaller than the SR-excitation rate--increases as a result of the decreasing emittance. At a wiggler length of approximately 75 m, the IBS rate exceeds the SR-excitation rate at the length coinciding with the transition between SR and IBS-dominated regimes. Unlike the natural emittance case, in the IBS-dominated regime SR excitation increases with wiggler length as the optimal solution favors additional damping to combat the rising IBS growth rate.
\begin{figure}
    \centering
    \includegraphics[width=1\linewidth]{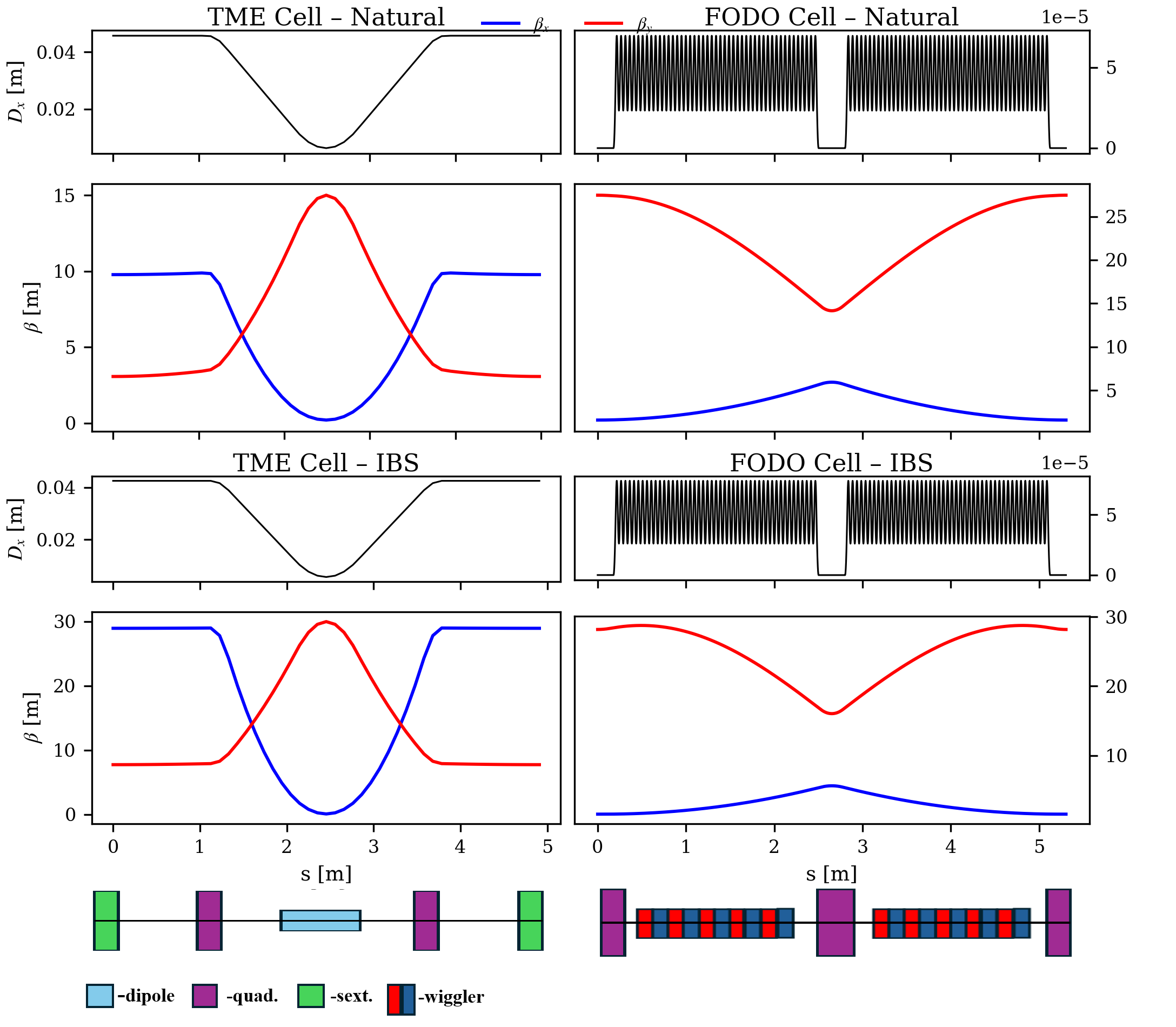}
    \caption{Example lattice functions for the case of a wiggler length of 360 m. The left column shows the TME cell for natural (top) and IBS cases (bottom) and the right shows the wiggler/FODO cell for natural (top) and IBS cases (bottom).}
    \label{fig:latticeFunctions}
\end{figure}

\begin{table}[t]
    \centering
    \label{tab:optimized_lattice}
    \begin{tabular}{lcc}
        \hline
        Optimization type & Natural & IBS \\
        \hline
        \multicolumn{3}{c}{\textit{Ring parameters}}\\
        \hline
        
        $\epsilon_x$ (nm) & 27 / 480 &  \quad 64 / 250 \\
        $\epsilon_y$ (nm) & 2.1 / 4.1 &  \quad 2.4 / 5.1 \\
        $(\sigma_E/E)$$\times10^{3}$ &  0.6 / 1.3 & \quad 0.7 / 1.3 \\
        $\sigma_z$ (mm) & 2 / 3.4 & \quad 2 / 3.3 \\
        $\tau_x$ (ms) & 5.9 & 2.6 \\
        $\tau_y$ (ms) & 5.9 & 2.6 \\
        $\tau_z$ (ms) & 3 & 1.3 \\
        \hline
        \multicolumn{3}{c}{\textit{TME cell parameters}}\\
        \hline
        $Q_x$ (per cell) & 0.46 & 0.49 \\
        $Q_y$ (per cell) & 0.17 & 0.07 \\
        $\xi_x$ (per cell) & -1.16 & -3.65 \\
        $\xi_y$ (per cell) &  -0.61 & -1.02 \\
        $L_{\rm cell}$ (m) & 4.42 & 4.45 \\
        $L_d$ (m) & \multicolumn{2}{c}{0.7} \\
         $N_d$  & \multicolumn{2}{c}{100} \\
        $\theta$ (mrad) & \multicolumn{2}{c}{62.98} \\
        $K_1^{Q}$ ($\mathrm{m}^{-2}$) & 3.98 & 4.24 \\
        $K_1^{B}$ ($\mathrm{m}^{-2}$) & -1.37 & -1.36 \\
        \hline
         \multicolumn{3}{c}{\textit{Wiggler Parameters}}\\
         \hline
         B-field (T) & 1.3 & 1.8 \\
         $\lambda_{wig}$ (cm) &  \multicolumn{2}{c}{4.9} \\
         $L_{\rm cell}$ (m) & \multicolumn{2}{c}{5.3} \\
         $N_{cell}$ & \multicolumn{2}{c}{74} \\
         $\langle \beta_x \rangle$ (m) & 3.0 & 2.9 \\
        \hline
    \end{tabular}
    \caption{Parameters of the natural-emittance and IBS-optimized lattices
    at a total wiggler length of 360~m. Beam parameters listed as pairs are
    given as natural (zero-charge) / IBS-equilibrium values, where the latter corresponds to a bunch charge of 1~nC.}
\end{table}

Example lattice functions of the TME and wiggler cells can be seen in Fig.~\ref{fig:latticeFunctions} for both the natural and IBS cases for a wiggler length of 360 m. For the TME cell the most significant difference between the lattice functions for both cases is larger values of $\beta_x$ and $\beta_y$ throughout the cell which results in a larger transverse beam size in the cell reducing the beam density. However, the larger beta functions necessarily detune from the optimal natural emittance value. This detuning is observed in both the SR and IBS-dominated regimes and the MOGA typically selects a value of $F\approx 4 F_{TME}$. The TME cell optimized for IBS requires stronger defocusing quads and larger beta functions resulting in an increase in the cell chromaticity.
In the wiggler cells, where SR excitation is proportional to $\langle \beta_x \rangle$ (see Eq.~\ref{I5_wig}), an increase in the beta functions is much less prominent. This behavior can be understood by noting that in the wiggler cells, SR emission generates emittance growth nearly uniformly over the full cell length, whereas in the TME cell SR emission is confined to the dipole region. As a result, detuning the lattice functions away from the natural-emittance optimum is more costly in the wiggler cells. In contrast, in the TME cell the SR-induced growth is localized, so deviation from the natural-emittance-optimal settings is less penalized. Because IBS acts along the entire cell, increasing the beta functions reduces the IBS rates everywhere, making detuning the TME cells a more effective strategy for lowering the charge-dependent emittance.

\begin{figure}
    \centering
    \includegraphics[width=1\linewidth]{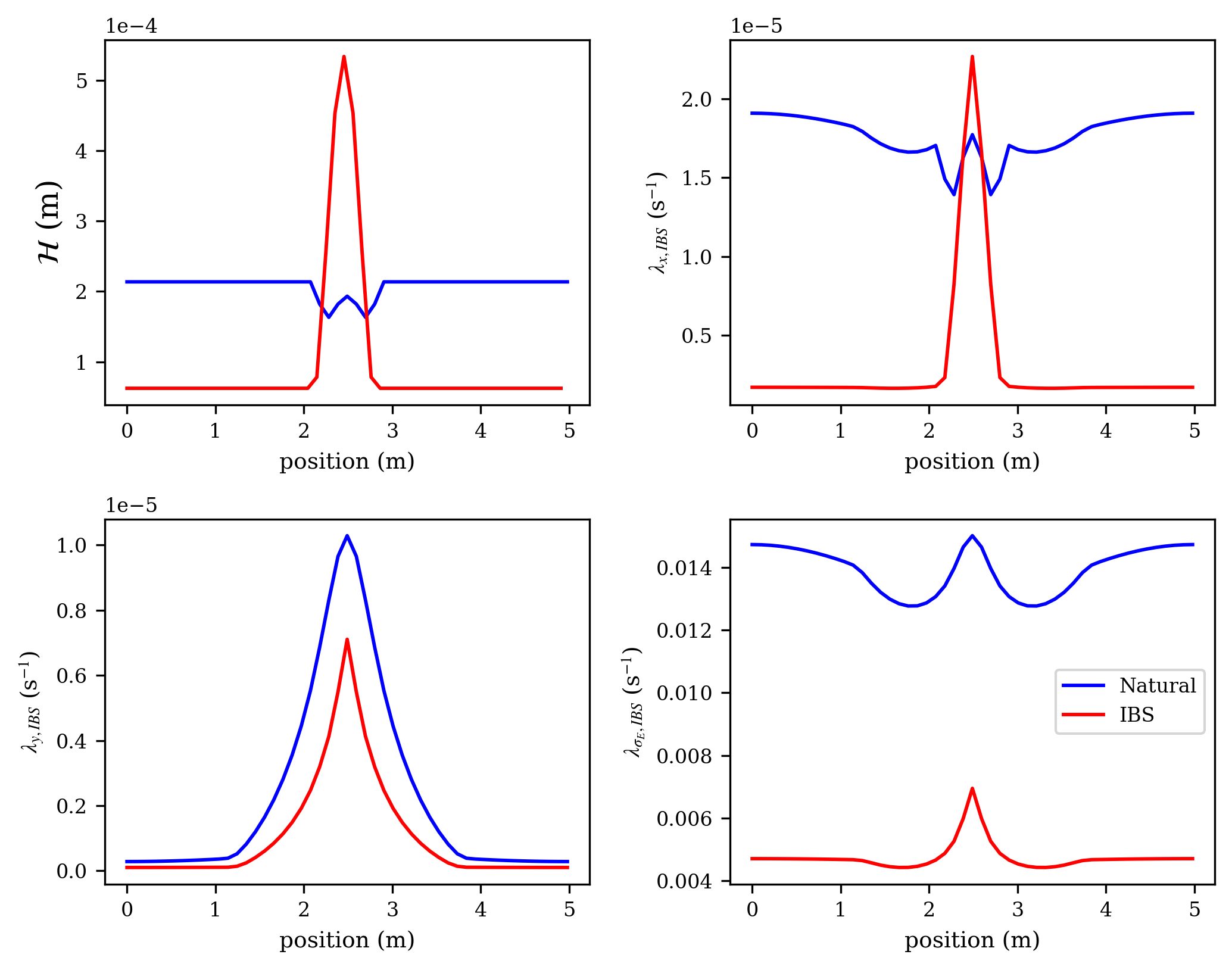}
    \caption{A comparison of the dispersion invariant $\mathcal{H}$ and the IBS rates for a TME cell optimized for natural emittance (blue) and IBS at 1 nC bunch charge (red).}
    \label{fig:IBSRates}
\end{figure}
The effect of detuning in the TME cell is illustrated in Fig.~\ref{fig:IBSRates}. The top left plot shows $\mathcal{H}$ which is minimized for the natural emittance case (blue curve) in the dipole region of the TME cell (refer to Fig.~\ref{fig:latticeFunctions} for cell layout). The IBS case shown in red is significantly larger in the dipole region. The other three plots show the IBS growth rates in the horizontal, vertical, and longitudinal directions. Although the MOGA optimization considered only the horizontal emittance, it can be seen that the IBS rates are reduced in all three planes which is the result of the larger beam size throughout the cell.
\subsection{Expected Performance}
For a total wiggler length of 360~m, the expected parameters of the $C^3$ damping ring at a beam energy of 1.54~GeV are shown in Table~\ref{tab:optimized_lattice}. For comparison, we also show the parameters of a ring with the same total wiggler length optimized for minimum natural emittance. For both lattices, the beam parameters are given for both the zero-charge (natural) and IBS-equilibrium cases. For the natural-emittance optimized lattice, IBS increases the horizontal emittance from 27 to 480~nm, an increase of approximately a factor of 17. In contrast, for the IBS-optimized lattice, the horizontal emittance increases from 64 to 250~nm, corresponding to a factor of only 3.9. The vertical emittance increases by approximately a factor of two due to  IBS for both lattices. This behavior results from two effects: (i) the zero-charge vertical emittance, set by the opening angle of SR, is larger because of the increase in $\langle\beta_y\rangle$ in the TME dipoles, and (ii) the smaller horizontal emittance of the IBS-optimized lattice leads to faster overall IBS growth rates. An important limitation of this study is that vertical dispersion is not included; consequently, the vertical emittance is determined solely by the direct IBS contribution and SR excitation from the finite emission angle.
\section{Additional Considerations}
\subsection{Polarization Preservation}
In this work, the beam energy is treated as a continuous variable subject only to the constraint imposed by the required damping time. Since $C^3$ will use polarized electrons, the energy should be chosen so that the spin tune is a half-integer~\cite{emma2001systematic}
\begin{equation}
a \gamma=n+1/2
\end{equation}
where $a\equiv(g-2)/2\approx 1.16\times10^{-3}$ is the anomalous magnetic moment and $n=0,1,2,\ldots$ is an integer. Thus, the allowed energies are in discrete jumps of approximately 440 MeV and an energy of 1.54 GeV is selected for the $C^3$ damping ring.
\subsection{Alignment Tolerances}
Using the analytic treatment developed in~\cite{Raubenheimer1991}, we estimate the normalized vertical emittance growth due to vertical dispersion introduced from vertical misalignment of the quadrupoles and bend magnets to be approximately 0.5 nm assuming an RMS alignment error of 20 $\mu$m for the 360 m wiggler case optimized for IBS. The natural emittance case had approximately a factor of 3 less emittance growth. The sensitivity found here is similar to what is reported for CLIC~\cite{Wootton2011}. Vertical dispersion introduced by quadrupole rotations resulted in negligibly small vertical emittance growth for errors of order 10 $\mu$rad. Emittance growth due to betatron coupling and sextupole alignment was not accounted in this work.
\section{Conclusion}
In this work the design of a DR was framed as a constrained optimization problem. A MOGA was used to map the trade-off between horizontal emittance and total wiggler length. MOGA results were checked against an analytic model for the natural emittance case. For the case of IBS, we identified two regimes: one where emittance is dominated by SR and the optimal wiggler B-field and beam energy match the natural case and an IBS-dominated regime where additional damping from a stronger wiggler field is beneficial. This framework was used to identify the major parameters for the $C^3$ linear-collider DR.
\section{Acknowledgments}
M.B. Andorf is grateful to A. Bartnik and C. Gulliford for help with TAO and Xopt. The authors thank M. Breidenbach for useful discussion. The work of the authors is supported by the U.S. Department of Energy under Contract No. DE-AC02-76SF00515.
\appendix
\section{}
\label{Appendix A}
Here we derive the criteria for when the inclusion of a damping wiggler results in a decrease in the horizontal emittance of the beam. The total horizontal emittance of a race-track layout ring with bending magnets in the arcs and damping wigglers in the dispersion free straight sections is given by
\begin{equation}
\epsilon=C_q\gamma^3\frac{I_{5,ring}+I_{5,wig}}{I_{2,ring}+I_{2,wig}}
\end{equation}
where we have assumed $j_x\approx 1$.  We first consider the case in which the energy of the ring remains fixed. Since both $I_{2,wig}$ and $I_{5,wig}$ are proportional to $L_{wig}$ we rewrite the above equation as
\begin{equation}
    f(x)=\frac{A+Bx}{C+Dx}
\end{equation}
where $A=I_{5,ring}$, $B=I_{5,wig}/L_{wig}$, $C=I_{2,ring}$, $D=I_{2,wig}/L_{wig}$ , $x=L_{wig}$ and we momentarily ignore the constant $C_q \gamma^3$ since it is not relevant for the argument. It is easy to show that for  $f'(x) < 0$, $AD>BC$. Multiplying both sides of the inequality by $L_{wig}$, and rearranging terms yields
\begin{equation}
\frac{I_{5,ring}}{I_{2,ring}}>\frac{I_{5,wig}}{I_{2,wig}}
\end{equation}
from which it follows
\begin{equation}
\epsilon_{wig}<\epsilon_{ring}.
\end{equation}
Thus, if the emittance of the wiggler (which is independent of $L_{wig}$
) evaluated as a stand-alone entity is less than the emittance of the ring without a damping wiggler, including a wiggler of any length will decrease the total emittance which will asymptotically approach  the value
\begin{equation}
\epsilon_{wig}=C_q\gamma^3\frac{I_{5,wig}}{I_{2,wig}}=C_q\gamma^3\frac{\pi^2\lambda_{wig}^2\langle\beta\rangle}{2^9\rho_{wig}^3}
\label{em_wig}
\end{equation}
for an infinitely long wiggler. 
In the case of the damping ring (not accounting for IBS), the beam energy should be set to the minimum energy given in Eq.~\ref{Emin}, in which case the total emittance can be expressed as
\begin{equation}
\epsilon=\mathcal{M}\frac{I_{5,ring}+I_{5,wig}}{(I_{2,ring}+I_{2,wig})^2}
\end{equation}
where $\mathcal{M}=\frac{C_q}{C_\gamma}\frac{4\pi L_{train}N_{\tau}f}{c (m_ec^2)^3}$ is a constant.  With similar substitutions as before we now have
\begin{equation}
f(x)=\frac{A+Bx}{(C+Dx)^2}.
\end{equation}
The derivative of the above function is 
\begin{equation}
f'(x)=\frac{CB-2AD-xBD}{(C+Dx)^3}.
\end{equation}
For any value $x > 0$ $f'(x)<f'(0)$. For any value of wiggler length to decrease the emittance the condition becomes $2AD>BC$ or 
\begin{equation}
\epsilon_{wig}<2\epsilon_{ring}
\end{equation}
where the factor of 2 is a direct consequence of the beam energy lowering with increased $I_2$. Note that for the case when $2AD<BC$ ($\epsilon_{wig}>2\epsilon_{ring}$) , $f(x)$ has a local maximum at $x=\frac{CB-2AD}{BD}$ but asymptotically approaches 0 as $x$ approaches infinity. Therefore, in this case, as the wiggler length is increased from 0 the emittance will initially grow before peaking and then reaching zero. This behavior is again a direct consequence of the energy decreasing with increasing $I_2$. In practice, as $\gamma$ decreases, IBS growth rates will dominate and the emittance is not well described by the above equations.
\section{}
\label{Appendix B}
Intrabeam Scattering (IBS) was implemented using the Completely Integrated Modified Piwinski approach~\cite{KUBO_CIMP}~\cite{Wolski:2009ibs}.
The growth rates are given by
\begin{equation}
\begin{aligned}
    \frac{1}{\tau_{x,IBS}}&=K_{IBS}\bigg\langle\frac{\mathcal H_x\sigma_H^2}{\epsilon_x}\bigg(\frac{1}{a}g\big(\frac{b}{a}\big)+\frac{1}{b}g\big(\frac{a}{b}\big)-ag\big(\frac{b}{a}\big)\bigg)\bigg\rangle\\
    \frac{1}{\tau_{y,IBS}}&=K_{IBS}\bigg\langle\frac{\mathcal H_y\sigma_H^2}{\epsilon_y}\bigg(\frac{1}{a}g\big(\frac{b}{a}\big)+\frac{1}{b}g\big(\frac{a}{b}\big)-bg\big(\frac{a}{b}\big)\bigg)\bigg\rangle\\
    \frac{1}{\tau_{p,IBS}}&=K_{IBS}\bigg\langle\frac{\sigma_H^2}{\sigma_p^2}\bigg(\frac{1}{a}g\big(\frac{b}{a}\big)+\frac{1}{b}g\big(\frac{a}{b}\big)\bigg)\bigg\rangle
\end{aligned}
\end{equation}
with 
\begin{equation}
\begin{aligned}
\frac{1}{\sigma_H^2}&=\frac{1}{\sigma_p^2}+\frac{\mathcal{H}_x}{\epsilon_x}+\frac{\mathcal{H}_y}{\epsilon_y}\\
a&=\frac{\sigma_H}{\gamma}\sqrt{\frac{\beta_x}{\epsilon_x}}\\
b&=\frac{\sigma_H}{\gamma}\sqrt{\frac{\beta_y}{\epsilon_y}}
\end{aligned}
\end{equation}
where the angled brackets denote averaging over the ring. The function $g(\omega)$ is given by
\begin{equation}
g(\omega)=\sqrt{\frac{\pi}{\omega}}\bigg[P_{-1/2}^0\bigg(\frac{\omega^2+1}{2\omega}\bigg)\pm \frac{3}{2}P_{-1/2}^{-1}\bigg(\frac{\omega^2+1}{2\omega}\bigg)\bigg]
\end{equation}
where $P_\nu^{-\mu}$ are the type 3 associated Legendre polynomials. The plus sign is used for $\omega \geq 1$ and the minus sign for $\omega \leq 1$. The evaluation of the associated Legendre functions was performed using the \texttt{mpmath} Python package~\cite{mpmath}. Numerical values were cross-checked in the table provided in~\cite{mtingwa1987intrabeam}.

To compute the average rates, the lattice functions were sampled at 10 cm intervals which is small compared to their rate of change. Note that we assume no alignment or coupling errors and therefore $\mathcal{H}_y=0$. The scale of the growth rates are given by $K_{IBS}$
\begin{equation}
K_{IBS}=2\pi^{3/2}\ln\bigg(\frac{\gamma^2\sigma_y\epsilon_x}{r_e\beta_x}\bigg)\frac{r_e^2cN_o}{64\pi^2\gamma^4\epsilon_x\epsilon_y\sigma_p}
\end{equation}
which includes the so-called Coulomb Log factor with no tail cut. In the above, $N_o$ is the number of particles and $r_e\approx 2.818\times10^{-15}$ m is the classical electron radius. The growth rates depend on the beam parameters and an equilibrium is met when growth from IBS and SR excitation are canceled by SR damping. The evolution toward equilibrium is described by
\begin{equation}
\begin{aligned}
    \frac{d\epsilon_i}{dt}&=-(\epsilon_i-\epsilon_{i,o})\frac{2}{\tau_i}+\epsilon_i\frac{2}{\tau_{i,IBS}(\epsilon_x,\epsilon_y,\sigma_p)} \\
    \frac{d\sigma_p}{dt}&=-(\sigma_p-\sigma_{p,0})\frac{1}{\tau_z}+\sigma_p\frac{1}{\tau_{p,IBS}(\epsilon_x,\epsilon_y,\sigma_p)}
\end{aligned}
\end{equation}
\begin{figure}
    \centering
    \vspace{12pt}  
    \includegraphics[width=1\linewidth]{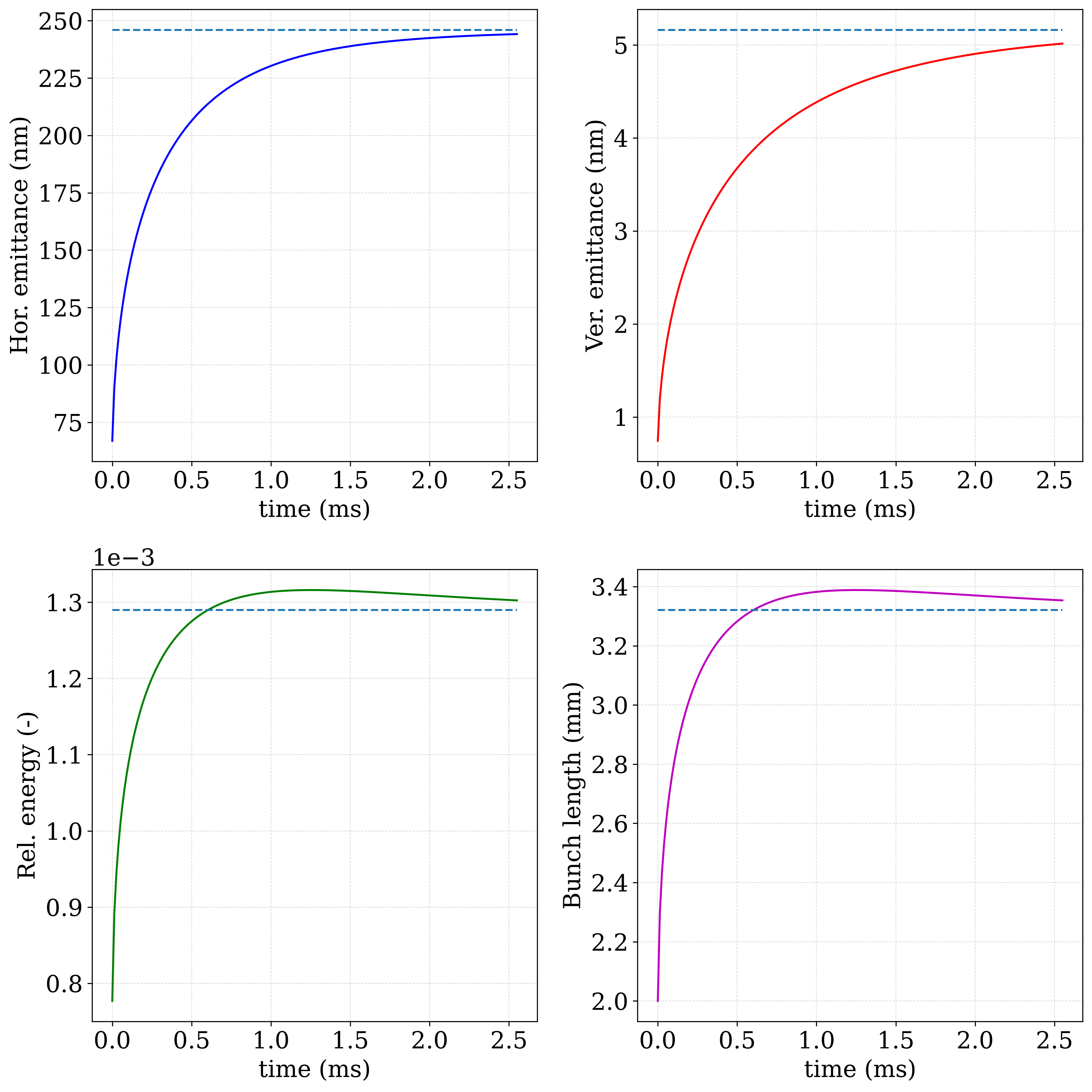}
    \caption{Example evolution of beam parameters moving toward equilibrium with IBS. The dashed line represents equilibrium found with the numeric solver.}
    \label{fig:IBS_evolution}
\end{figure}
where $i=x,y$ refers to the horizontal or vertical plane, $\epsilon_i$ and $\epsilon_{i,o}$ are the transverse geometric and natural geometric emittances while $\sigma_p$ and $\sigma_{p,o}$ are the energy and natural energy spreads, $\tau_i$ and $\tau_z$ are the SR damping times in the transverse and longitudinal planes due to SR and $\tau_{i,IBS}$  and $\tau_{p,IBS}$ are the growth times from IBS in the transverse and longitudinal planes. For the vertical plane, the natural geometric emittance is set by the radiation opening angle. To find the equilibrium emittance, the above equations are advanced numerically with a time step $\tau_{min}/20$ for several damping times where $\tau_{min}$ is the shortest SR damping time of the three planes. At each time step, the IBS growth rates are re-evaluated and the bunch length is updated from the instantaneous energy spread with the standard relation $\sigma_z=\alpha_pc\sigma_p/\omega_s$ where $\alpha_p$ is the momentum compaction and $\omega_s$ is the synchrotron tune. The selected initial conditions were $\epsilon_x(t=0)=\epsilon_{x,o}$, $\epsilon_{y}(t=0)=\epsilon_{x,o}/100$ and $\sigma_p(t=0)=\sigma_{p,o}$. The initial conditions were selected for numerical convenience and do not represent the beam injected into the DR.

Once the solutions have been advanced near equilibrium, a non-linear solver utilizing the modified Powell hybrid method,  via SciPy's implementation \cite{Virtanen2020}, was used to obtain the precise equilibrium value. Convergence testing was performed for different initial conditions, time step and total integration time. An example of the beam evolving from the chosen initial conditions toward equilibrium is shown in Fig.~\ref{fig:IBS_evolution}. Our IBS implementation was cross-checked against BMAD's \texttt{ibs\_ring} \cite{Bmad:IbsRing} and agreement was found to be within 2 $\%$ for all beam parameters. 
\bibliographystyle{unsrt}   
\bibliography{references}

\end{document}